\documentclass[12pt]{iopart}
\usepackage{iopams}
\usepackage{epsfig}   
\usepackage{graphics}
\usepackage[T1]{fontenc}

\newcommand{\av}[1]{\langle{#1}\rangle}

\begin{document} 

\title{Apparent and average acceleration of the Universe}

\author{Krzysztof Bolejko$^{1,2}$ and Lars Andersson$^{3,4}$}

\address{$^1$School of Physics, University of Melbourne, VIC 3010, Australia}
\address{$^2$Nicolaus Copernicus Astronomical Center, Bartycka 18, 00-716 Warsaw, Poland}
\address{$^3$Department of Mathematics, University of Miami, Coral Gables, FL 33124, USA}
\address{$^4$Albert Einstein Institute, Am M\"uhlenberg 1, D-14467 Golm, Germany}

\ead{\mailto{bolejko@camk.edu.pl}, \mailto{larsa@math.miami.edu}}

\begin{abstract}
In this paper we consider the relation between the volume deceleration parameter
obtained within the Buchert averaging scheme and the deceleration parameter
derived from the supernova observation. This work was motivated by recent
findings that showed that there are models which despite $\Lambda=0$ have 
volume deceleration parameter $q^{vol} < 0$. This opens the
possibility that backreaction and averaging effects may be used 
as an interesting alternative explanation to the dark
energy phenomenon.  

We have calculated $q^{vol}$ in some Lema\^itre--Tolman models. For those models
which are chosen to be realistic and which fit the supernova data, we find
that $q^{vol} > 0$, while those models which we have been able to find which
exhibit $q^{vol} < 0$ turn out to be unrealistic. This indicates that 
care must be exercised in relating the deceleration parameter 
to observations. 
\end{abstract}



\section{Introduction}
Accelerated expansion, modeled by a positive cosmological constant, 
is an essential element of the current standard cosmological model 
of the Universe. The accelerated expansion was originally motivated by 
supernova observations \cite{sn} and is supported by many other
types of cosmological observations. Observational data is, in modern cosmology, 
analyzed almost exclusively within the 
framework of homogeneous and isotropic Friedmann models \cite{grs}. This analysis
leads to the Concordance model, which 
provides a remarkably precise fit to cosmological observations.
In this situation, if the 
Ehlers-Geren-Sachs theorem \cite{EGS} and `almost EGS
theorem' \cite{SME} are invoked\footnote{These theorems imply that if anisotropies in the cosmic microwave background radiation are small for all fundamental observers then the Universe is locally almost spatially homogeneous and isotropic. However, as shown in \cite{NUWL99} the almost Robertson--Walker geometry also requires the smallness of the Weyl curvature.}, 
then it seems that an assumption of large scale homogeneity of
the Universe can be justified. This on the other hand implies that the Universe must be filled
with dark energy which currently drives the acceleration of the Universe.

However the Concordance model is not the only one which can fit cosmological 
observations. Anti-Copernican inhomogeneous models which assume
the existence of a local Gpc scale void also fit cosmological observations
\cite{ide} (see \cite{C07} for a review).
Moreover, on small and medium scales our Universe is not
homogeneous. Therefore, one may ask 
whether Friedmann models can describe our
Universe correctly. In particular, it is important to ask 
what is the best way to fit a homogeneous model to a realistic
and inhomogeneous Universe. This problem, known as the fitting problem,
was considered by Ellis and Stoeger \cite{ES87}.
In considering the fitting problem, it becomes apparent that 
a homogeneous model fitted to inhomogeneous data can evolve quite differently 
from  the real Universe.
The difference between evolution of homogeneous
models and an inhomogeneous Universe is caused by backreaction effects, due
to the nonlinearity of the Einstein equation. 
Unfortunately, in the standard approach, the backreaction is rarely
taken into account -- in most cases when
modelling our Universe on a local scale Newtonian 
mechanics is employed and on large scales the Friedmann
equations (or linear perturbations of Friedmann background) are used \cite{P93}. 
Such an approach to cosmology is often encouraged by the ``no--go'' theorem
which states that the Universe can be very accurately described
by the conformal Newtonian metric perturbed about a spatially flat background,
even if $\delta \rho / \rho \gg 0$. In such a case the backreaction 
is negligible \cite{IW06,KAF06}.
However, the results obtained by van Elst and Ellis \cite{EE98} and
recently by Kolb, Marra and Matarrese \cite{KMM08}
show that the application of ``no-go'' theorem is limited.
Therefore, one should be aware that in the absence of an analysis of the
backreaction and other effects caused by inhomogeneities in the universe, 
there remains the possibility that the observed accelerated expansion of the Universe is
only apparent \cite{El08}. 
The direct study of the dynamical effects of inhomogeneities is difficult.
Due to the nonlinearity of the Einstein equations, the solution of the Einstein equations for the homogeneous matter distribution leads in principle to a different description  
of the Universe than an average of a inhomogeneous solution to the exact
Einstein equations (even though inhomogeneities when averaged  
over a sufficiently large scale might tend to be zero).

Neither the analysis of the evolution of a
general matter distribution nor the numerical evolution of cosmological models
employing the full Einstein equations are available at the level of
detail which would make them useful for this problem. 
There are currently several different approaches which attempt to take 
backreaction effects into account. 
One approach is based on exact solutions -- see for example
\cite{ha}. 
Another, and more popular approach is based on averaging.

In the averaging approach to backreaction, one considers a solution to 
the Einstein equations for a general matter distribution and 
then an average of various observable quantities is taken. 
If a simple volume average is considered then such an attempt leads
to the Buchert equations \cite{B00}. The Buchert equations are very
similar to the Friedmann equations except for the backreaction term which is
in general nonvanishing, if inhomogeneities 
are present.
For a review on backreaction and the
Buchert averaging scheme the reader is referred to
\cite{SR2006,B08}.  
Within this framework and using spherically symmetric
inhomogeneous models  
Nambu and Tanimoto \cite{NT}, Paranjape and Singh \cite{PS}, Kai, Kozaki, Nakao, Nambu, and Yoo
\cite{KKNNY}, Chuang, Gu, and Hwang \cite{CGH},
provided explicit examples that one can obtain negative values of the volume deceleration parameter
even if $\Lambda = 0$.
Another interesting example was presented by R\"as\"anen
\cite{SR2006,R06b} where 
it was shown that the total volume deceleration parameter 
of two isolated and locally decelerating regions
can also be negative.

There are however important ambiguities in the application of an averaging
procedure. The average itself not
only depends on a choice of volume but also on a choice of time slicing. 
This is very crucial in
cosmology. Once inhomogeneities are present the age of the Universe is not
everywhere the same. Namely, the big bang in inhomogeneous models is not a
single event, so the average taken over a hypersurface of constant cosmic
time $t$ is different from the average taken over a hypersurface of constant
age of the Universe $t - t_B$ \cite{SR2004b}.
Moreover, the results of the averaging procedure vary 
if the discrepancy between the average
cosmic time and the local time is introduced (the local time is the time
which is measured by local clocks; the cosmic time is the time which appears
in the averaged homogeneous model). This phenomenon was studied by Wiltshire
\cite{Wa}, and has been used in an ambitious alternative concordance
model. The model proposed by Wiltshire introduces some additional assumptions which allow 
to some extent a comparison of averaged quantities with observations.
Such a comparison shows quite good agreement with observations, \cite{LNW08}. Thus, while serious
fundamental questions remain concerning Wiltshire's approach, it is another
example of an approach where one does not need  dark energy
to fit cosmological observations.

The averaging procedure is also gauge-dependent.
For example using different gauge one can obtain that the backreaction mimics
not dark energy but dark matter \cite{KKM}.
The averaging schemes, therefore, in the literature have been criticized, and their
inherent ambiguities (and in some cases obscurity) have been discussed,
cf. e.g. \cite{IW06}.  A key point is that it is far from obvious if the
average quantities, such as the acceleration of the averaged universe are 
really the quantities which are measured in astronomical observations. 
In particular, an operational analysis is to a large extent lacking in the
discussions of averaging. Thus, it is important to test the
averaging procedures with the exact and inhomogeneous solutions of the
Einstein equations. Within exact models each quantity can easily be
calculated and then compared with its averaged counterpart.  This paper aims
to perform such an analysis within the Lema\^itre--Tolman model.

The structure of this paper is as follows. Buchert's averaging procedure
is presented in section \ref{BSsec}, and some background on the  
Lema\^itre--Tolman model is given in section \ref{LTsec}. The
volume and distance deceleration parameters are introduced
in section \ref{DPsec}. 
Finally, in section  \ref{QAsec}, we discuss the relation between the 
deceleration parameters, supernova observations and models of cosmic
structures.

\section{The Buchert scheme}\label{BSsec}

If the averaging procedure is applied to the Einstein equations, then for
irrotational and pressureless matter the following equations are obtained
\cite{B00}

\begin{eqnarray}
&& 3 \frac{\ddot{a}}{a} = - 4 \pi G \av{\rho} + \mathcal{Q},
\label{bucherteq1} \\ &&  3 \frac{\dot{a}^2}{a^2} = 8 \pi G \av{ \rho}	-
\frac{1}{2} \av{ \mathcal{R} } - \frac{1}{2} \mathcal{Q}, \label{bucherteq2} \\
&&  \mathcal{Q} \equiv \frac{2}{3}\left( \av{{\Theta^2}} - \av{ \Theta }^2 \right)
- 2 \av{ \sigma^2},
\label{qdef}
\end{eqnarray}
where $\av{ \mathcal{R} }$ is an average of the spacial Ricci scalar $^{(3)}
\mathcal{R}$, $\Theta$ is the scalar of expansion, $\sigma$ is the shear
scalar, and $\av{\ }$ is the volume average over the hypersurface of constant time:
$\av{A} = ( \int d^3x \sqrt{-h} )^{-1} \int d^3x \sqrt{-h} A $. The scale factor $a$ is
defined as follows:

\begin{equation}
a = (V/V_0)^{1/3},
\label{aave}
\end{equation}
where V$_0$ is an initial volume.

Equations (\ref{bucherteq1}) and (\ref{bucherteq2}) are very similar to the
Friedmann equations, where Q=0, and $\rho$ and $\mathcal{R}$ depend on time
only.  In fact, they are 
kinematically
equivalent with a Friedmann model that has an additional scalar field    
source \cite{BLA06}. 
However the Buchert equation do not form a closed system.
To close these equation one has to introduce some further assumptions \cite{B00}.
As can be seen from  (\ref{qdef}) if the dispersion of expansion is
large, Q can be large as well and one can get acceleration ($\ddot{a}>0$)
without employing the cosmological constant.

\section{The Lema\^itre--Tolman model}\label{LTsec}

The Lema\^itre--Tolman model\footnote{The pressure free and irrotational solution 
of the Einstein equations for spherically symmetric space-time is often
called the Tolman, Tolman--Bondi, or Lema\^itre--Tolman--Bondi model.
However, it is more justified to refer to this solution
as to the Lema\^itre--Tolman model (cf. \cite{K97}).}
 \cite{LT} is a spherical symmetric,
pressure free and irrotational solution of the Einstein equations.  Its
metric is of the following form

\begin{equation}
{\rm d}s^2 =  c^2{\rm d}t^2 - \frac{R'^2(r,t)}{1 + 2 E(r)}\ {\rm d}r^2 -
R^2(t,r) {\rm d} \Omega^2, \label{ds2}
\end{equation}
where $ {\rm d} \Omega^2 = {\rm d}\theta^2 + \sin^2 \theta {\rm
d}\phi^2$. Because of the signature $(+, -, -, -)$, the $E(r)$ function must
obey $E(r) \ge - 1/2.$ Prime $'$ denotes $\partial_r$.

The Einstein equations reduce, in $\Lambda=0$ case, to the following two
\begin{equation}\label{den}
\kappa \rho(r,t) c^2 = \frac{2M'(r)}{R^2(r,t) R'(r,t)},
\end{equation}
\begin{equation}\label{vel}
\frac{1}{c^2}\dot{R}^2(r,t) = 2E(r) + \frac{2M(r)}{R(r,t)},
\end{equation}
\noindent where $M(r)$ is another arbitrary function and $\kappa = 8 \pi
G/c^4$. Dot $\dot{}$ denotes $\partial_t$.

When $R' = 0$ and $M' \ne 0$, the density becomes infinite. This happens at
shell crossings. This is an additional singularity to the Big Bang that
occurs at $R = 0, M' \neq 0$.  By setting the initial conditions
appropriately the shell crossing singularity can be avoided (see
\cite{HL1985} for detail discussion).

Equation (\ref{vel}) can be solved by simple integration:

\begin{equation}\label{evo}
\int\limits_0^R\frac{d\tilde{R}}{\sqrt{2E + \frac{2M}{\tilde{R}}}} = c \left[t- t_B(r)\right],
\end{equation}
where $t_B$ appears as an integration constant and is an arbitrary function
of $r$. This means that the big bang is not a single event as in the
Friedmann models, but occurs at different times at different distances from
the origin.

The scalar of the expansion is equal to

\begin{equation}
\Theta =  \frac{\dot{R}'}{R'} + 2 \frac{\dot{R}}{R}.
\label{theta}
\end{equation}

The shear tensor is of the following form:

\begin{equation}
 \sigma^\alpha{}_\beta = \frac{1}{3} \left( \frac{\dot{R}'}{R'} -
  \frac{\dot{R}}{R} \right) {\rm diag} (0, 2, -1, -1),
\end{equation}
thus $\sigma^2 \equiv (1/2) \sigma_{\alpha \beta} \sigma^{\alpha \beta}
=(1/3) (\dot{R}'/R' - \dot{R}/R)^2$.

The spacial Ricci scalar in the Lema\^itre--Tolman is equal to

\begin{equation}
^{(3)} \mathcal{R} = - \frac{4}{R^2} \left( E + \frac{E' R}{R'} \right)
\end{equation}

\section{The apparent and average acceleration}\label{DPsec}

The deceleration parameter within the Friedmann models is defined as

\begin{equation}
q =  - \frac{\ddot{a}a}{\dot{a}^2},
\label{decparflrw}
\end{equation}
where $a$ is the scale factor.
By analogy we can define the deceleration parameter which is based on the
averaging scheme. Substituting (\ref{aave}) into (\ref{decparflrw}) and using
(\ref{bucherteq1}) and (\ref{bucherteq2}) we get

\begin{equation}
q^{vol} = - \frac{- 4 \pi G \av{ \rho } + \mathcal{Q}}{8 \pi G \av{ \rho }  -
\frac{1}{2} \av{ \mathcal{R} } - \frac{1}{2} \mathcal{Q}}.
\label{decparave}
\end{equation}
We refer to this deceleration parameter as the {\it volume deceleration parameter}, $q^{vol}$
since it is positive when the second derivative of volume is negative
and negative when the second derivative of volume is positive (and of sufficiently large value).

On the other hand one can introduce a deceleration parameter
defined relative to the distance. Within homogeneous models the
distance  to a given redshift is larger for accelerating models than for
decelerating ones.
Taylor
expanding the luminosity distance in the Friedmann model we obtain 
\begin{eqnarray}
&& D_L =	\left. \frac{ d D_L}{d z} \right|_{z=0} z + \frac{1}{2}
\left. \frac{ d^2 D_L}{d z^2} \right|_{z=0} z^2 + \mathcal{O}(z^3) \nonumber
\\ && = \frac{c}{H_0} z + \frac{c}{2 H_0} (1 - q) z^2 + \mathcal{O}(z^3).
\label{dlfl}
\end{eqnarray}
Employing a similar procedure in the case of 
the Lema\^itre--Tolman model we get 
\begin{equation}
D_L =  \frac{cR'}{\dot{R}'} z  + \frac{c}{2} \frac{R'}{\dot{R}'} \left( 1 +
\frac{R' \ddot{R}'}{\dot{R}'^2} + \frac{cR''}{R' \dot{R}'} -
\frac{c \dot{R}''}{\dot{R}'^2} \right) z^2 + \mathcal{O}(z^3),
\label{dllt}
\end{equation}
Thus by comparing (\ref{dllt}) with  (\ref{dlfl}), the Hubble and the
deceleration parameter in the Lema\^itre--Tolman model can be defined as
\begin{equation}
H^{dis}_0 = \frac{\dot{R}'}{R'}, \quad q^{dis}_0 =  - \frac{R' \ddot{R}'}{\dot{R}'^2} - \frac{c R''}{R' \dot{R}'} +
\frac{c \dot{R}''}{\dot{R}'^2}.
\label{qddf}
\end{equation}
The above quantities are defined at the origin ($r=0$). However, following the Partovi and Mashhoon \cite{PM84} we can extend the above quantities to any $r$. Then, the coefficients of Taylor expansion are

\begin{eqnarray}
 \frac{ d D_L}{d z}  = && 2 R + \dot{R} \frac{dt}{dz} + R' \frac{dr}{dz} \nonumber \\
\frac{ d^2 D_L}{d z^2} = && 2 R + 4 \dot{R} \frac{dt}{dz} + 4 R' \frac{dr}{dz} 
+ \ddot{R} \left(\frac{dt}{dz}\right)^2 + 2 \dot{R}' \frac{dt}{dz} \frac{dr}{dz} 
+ R'' \left(\frac{dr}{dz}\right)^2 \nonumber \\
&& + \dot{R} \frac{d^2t}{dz^2} + R' \frac{d^2r}{dz^2},
\end{eqnarray}
and we obtain

\begin{equation}
H^{dis} =  \left( \frac{ d D_L}{d z} \right)^{-1}, \quad q^{dis} =  1 -H^{dis} \frac{ d^2 D_L}{d z^2}.
\label{qdis}
\end{equation}
We refer to this deceleration parameter as the {\it distance deceleration parameter}.
Although of physical importance is the luminosity distance
and its ability of fitting the supernova data, the $q^{dis}$ is
of great usefulness. It allows us, 
without solving the geodesic equations, to easily check whether  
a considered model can be use to fit supernova data.
As we will see in the next section, models which fit supernova data have 
at least in some regions $q^{dis}<0$.

\section{Connection between deceleration parameter and observations}\label{QAsec}

\begin{figure}
\begin{center}
\includegraphics[scale=0.7]{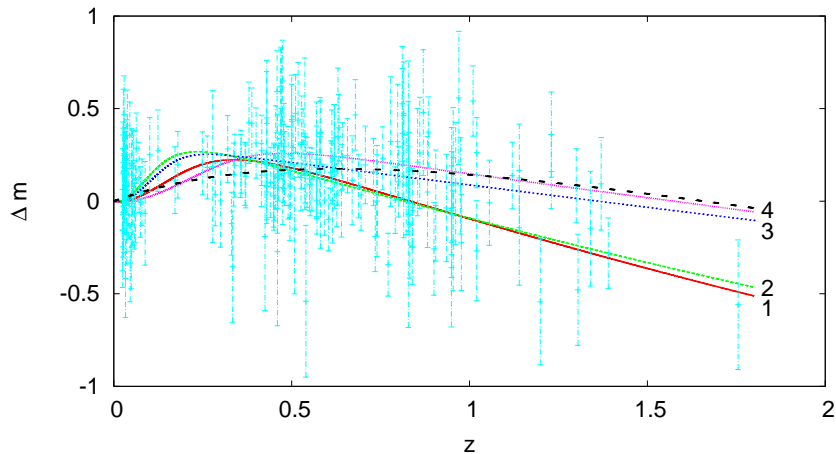}
\caption{The Residual Hubble diagram for models 1-4.  The black dashed line 
presents $\Delta {\rm m}$ for the $\Lambda$CDM model.}
\label{f1}
\end{center}
\end{figure}

\begin{figure}
\includegraphics[scale=0.65]{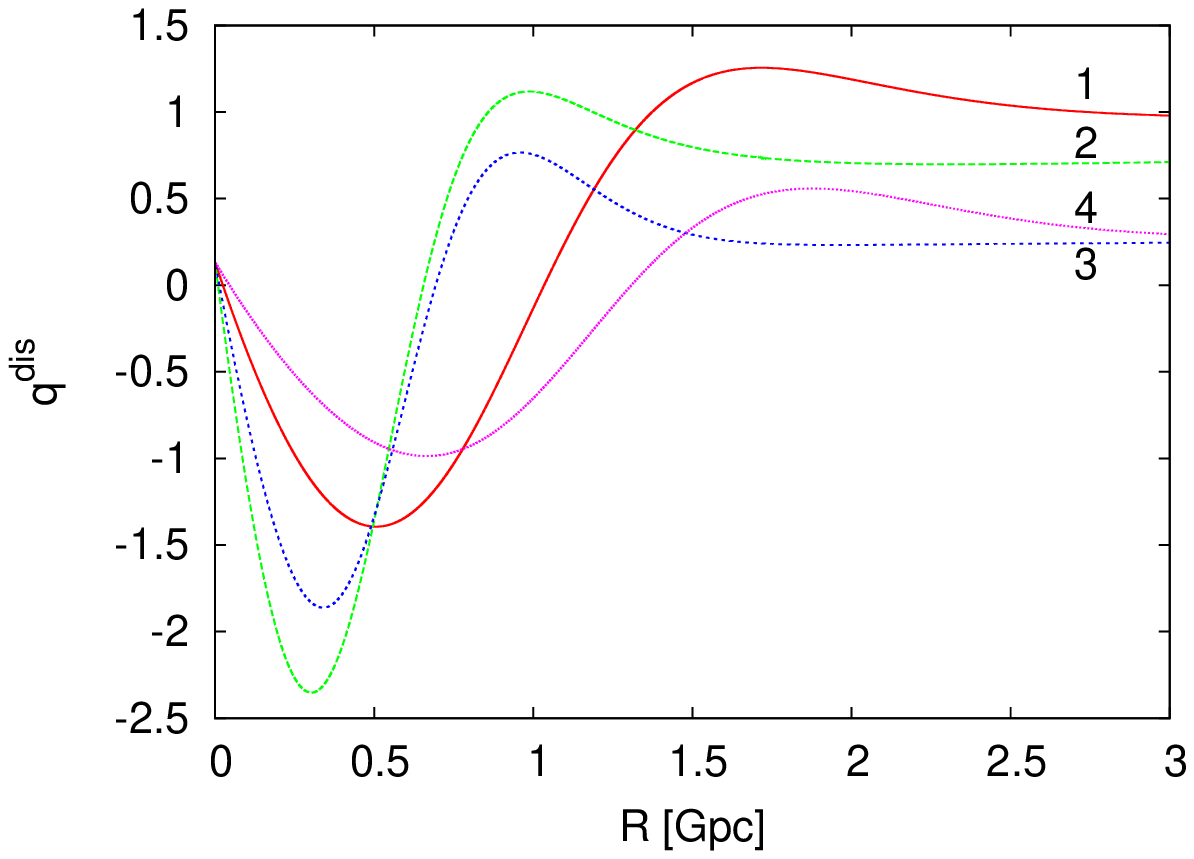} 
\includegraphics[scale=0.65]{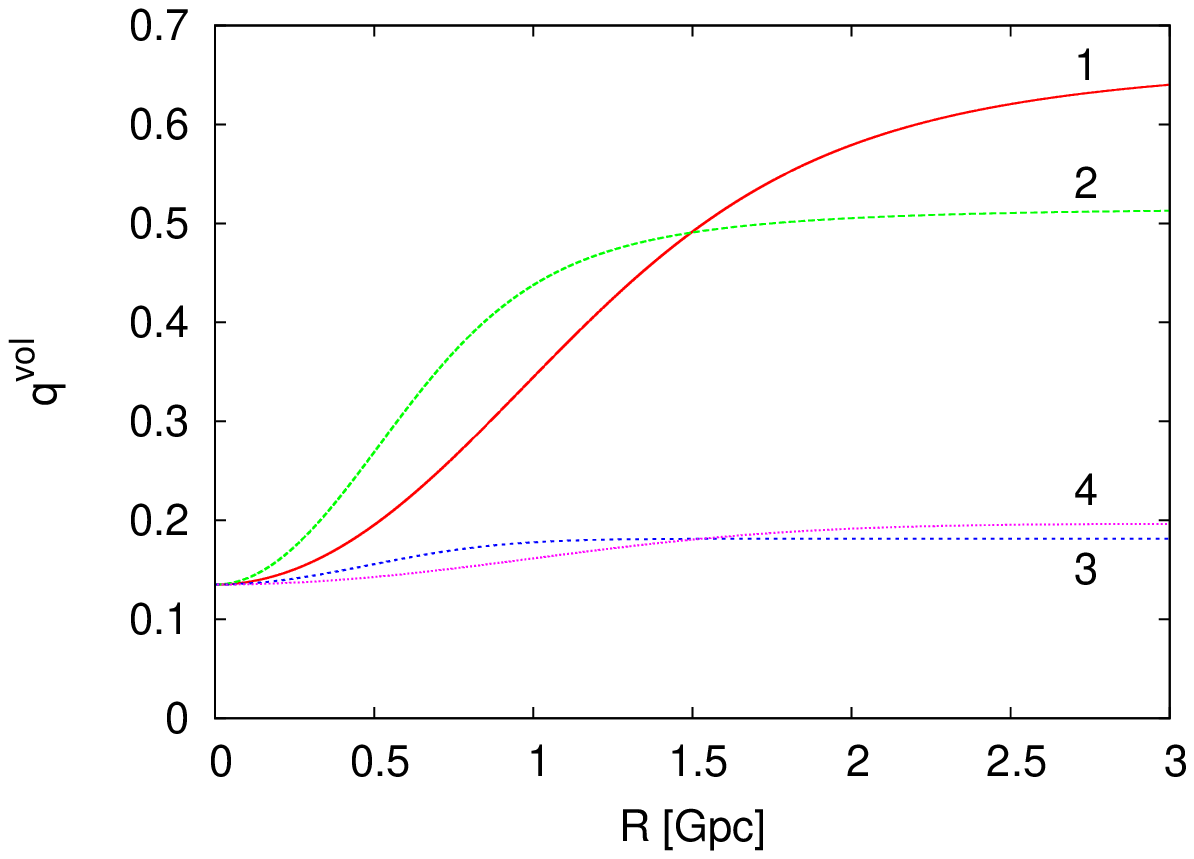}
\caption{The distance deceleration parameter (left panel) and the volume
deceleration parameter (right panel) for models 1-4.}
\label{f2}
\end{figure}

Let us first focus on supernova observations. There is already a considerable
literature on 
inhomogeneous models which are able to fit the supernova observations without the
cosmological constant
\cite{ide}.
We shall examine four such models in this section. 
For each of these models we shall calculate 
the volume and distance deceleration parameters
and compare with each other.
The four models to be considered present a very good fit to
supernova data. The supernova data consists of 182 supernovae from 
the Riess gold sample \cite{R07}. The $\chi^2$ test for models 1-4 is
respectively 183.6, 184.3, 164.7, and 178.5 (for comparison
the $\chi^2$ of fitting the $\Lambda$CDM model is 165.3).
The residual Hubble diagram for
these models is presented in figure \ref{f1}.
The deceleration parameters for models 1-4 are presented in figure
\ref{f2}. Left panel presents the distance deceleration parameter [as defined
by (\ref{qdis}) - where $dt/dz$ and $dr/dz$ were calculated for the radial geodesic]. 
The distance deceleration parameter is positive at the origin, but soon becomes negative.
Moreover, a very similar shape is obtained if instead 
$q^{dis}$ [as defined by (\ref{qdis})] $q^{dis}_0$ [as defined by (\ref{qddf})] is used.
Thus, $q^{dis}$ (or even $q^{dis}_0$, if treated as a function of $r$) can be regarded as a useful test to check if
a given model is able to fit supernova data.
However, the most significant is that the volume deceleration parameter
which is presented in the right panel of figure \ref{f2} is strictly positive.
Thus, the ability of reproducing the supernova data does not
require that the volume deceleration parameter is negative. This raises the question whether the average acceleration has 
any relation with the observed acceleration of the Universe; and if yes, are models with
average acceleration also able to fit supernova data?

\begin{figure}
\includegraphics[scale=0.65]{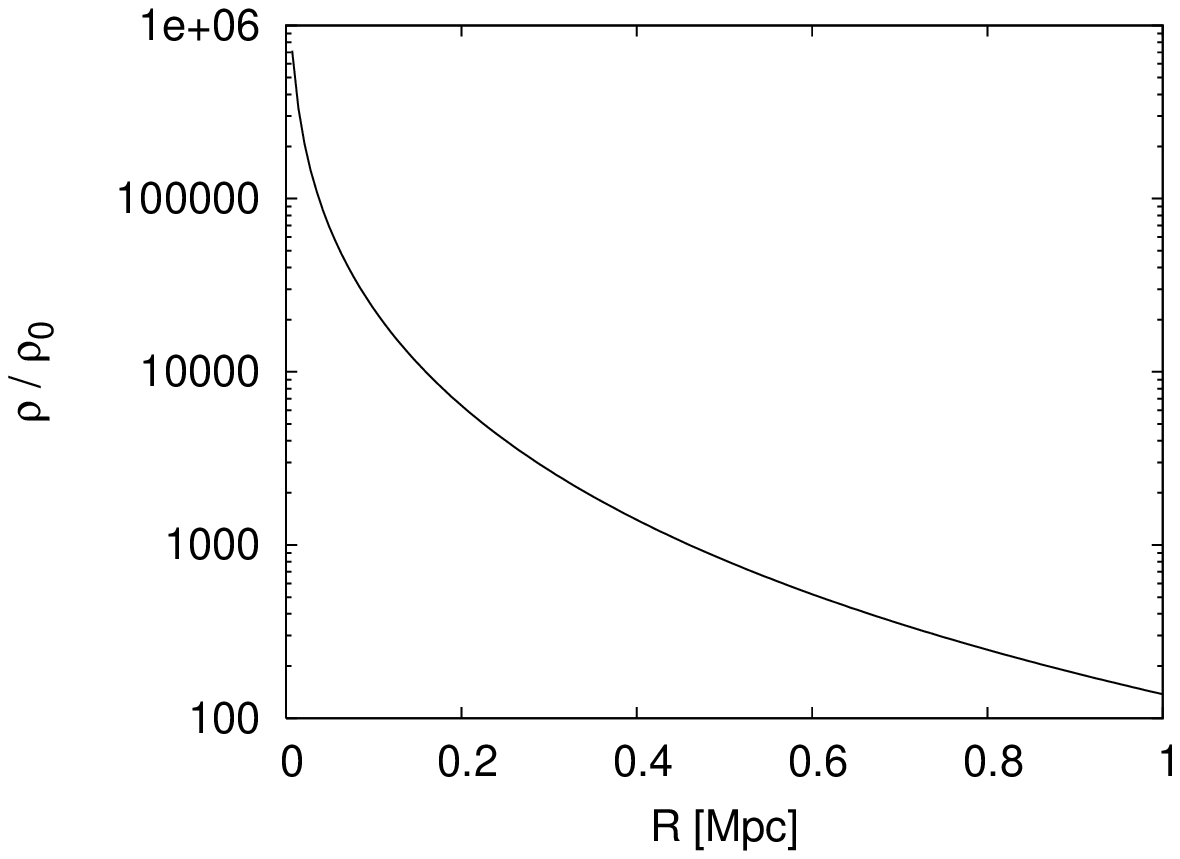} 
\includegraphics[scale=0.65]{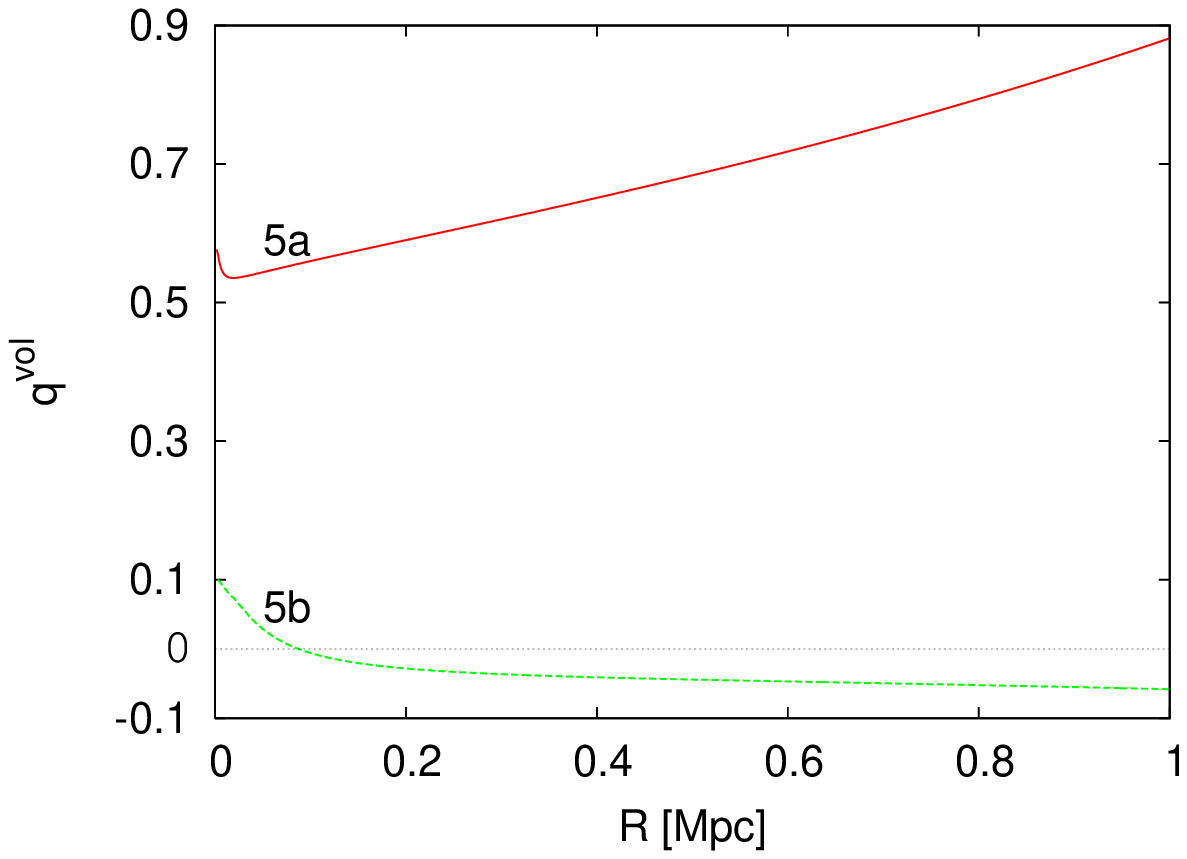}
\caption{The current density distribution (left panel) and deceleration
parameter (right panel) for model 5.}
\label{f3}
\end{figure}

Let us now focus on models of cosmic structures.  It was recently shown that
using a perturbative approach, backreaction cannot explain the apparent
acceleration \cite{pab}. 
However, because of large density
fluctuations within cosmic structures, results obtained in terms of the
perturbation framework might be questionable.  Moreover, 
in view of the fact 
that there are known examples of exact inhomogeneous models with negative
volume deceleration parameter and $\Lambda = 0$, it is worthwhile to check
if realistically evolving models of cosmic structures can have negative
values of deceleration parameter.
First, let us consider a model of galaxy clusters with the Navarro-Frenk-White
density distribution \cite{NFW} (left panel of figure \ref{f3}). 
Although, the NFW profile describes virialized systems\footnote{The Lema\^itre--Tolman model which evolve from smooth density profile at last scattering
to a high value profile like the NFW profile is always characterized 
by a collapse -- central region within this models are at the current instant collapsing. Thus such systems cannot be considered as virialized systems.}
the use of this profile will prove to be very instructive.
The average deceleration parameter $q^{vol}$ for model 5 is presented in the right panel of
figure \ref{f3}.  As can be seen in this case the deceleration parameter is
positive (curve 5a). However, it is possible to modify this model so 
that the 
$q^{vol}$ becomes negative -- curve 5b in the right panel of figure \ref{f3}.  
This was obtained by choosing the $E$ function which is of large positive value
(for details see Appendix). However,  after such a 
modification this model becomes unrealistic.  Specifically, 
the age of the Universe in this model becomes unrealistically small. The bang time
function $t_B$ in this model is of large amplitude, around $11.44 \times
10^9$ y. This means that the actual age of the Universe in this model is
approximately a few hundreds of thousand years.

\begin{figure}
\includegraphics[scale=0.65]{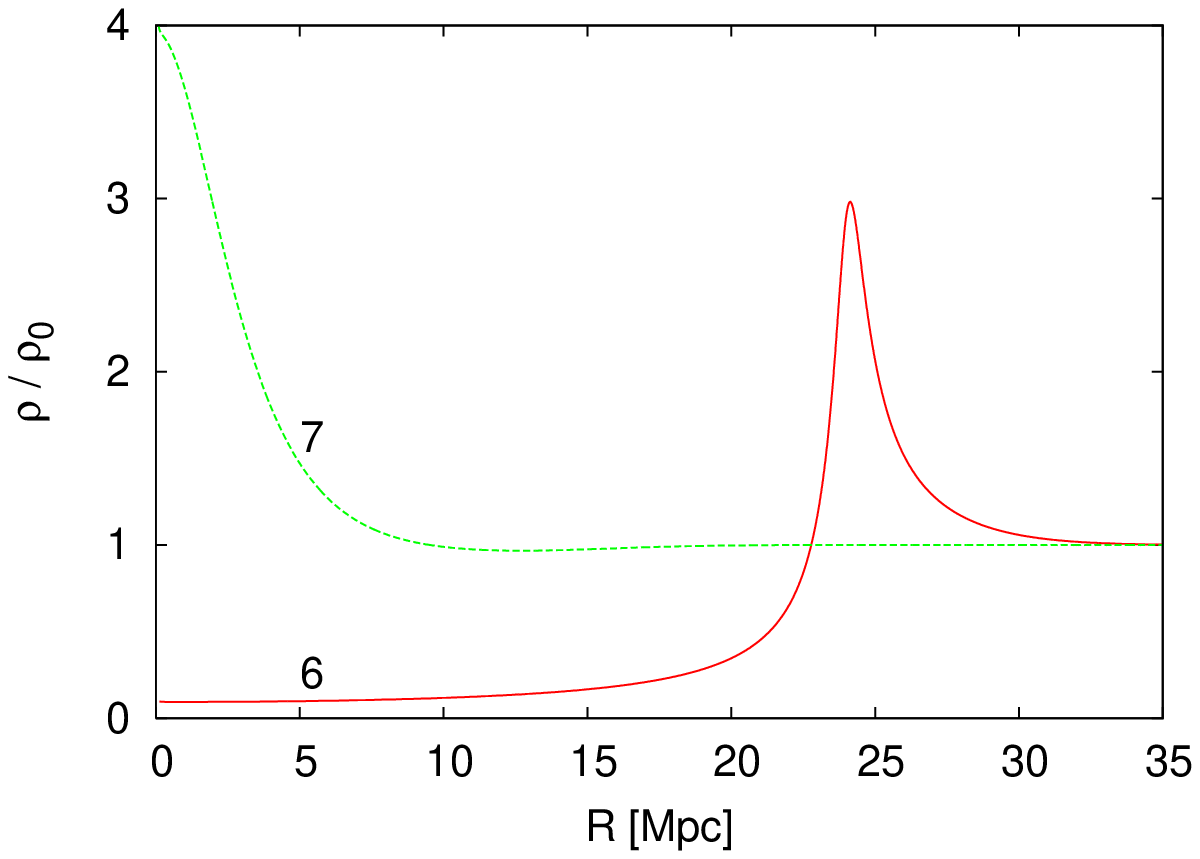}
 \includegraphics[scale=0.65]{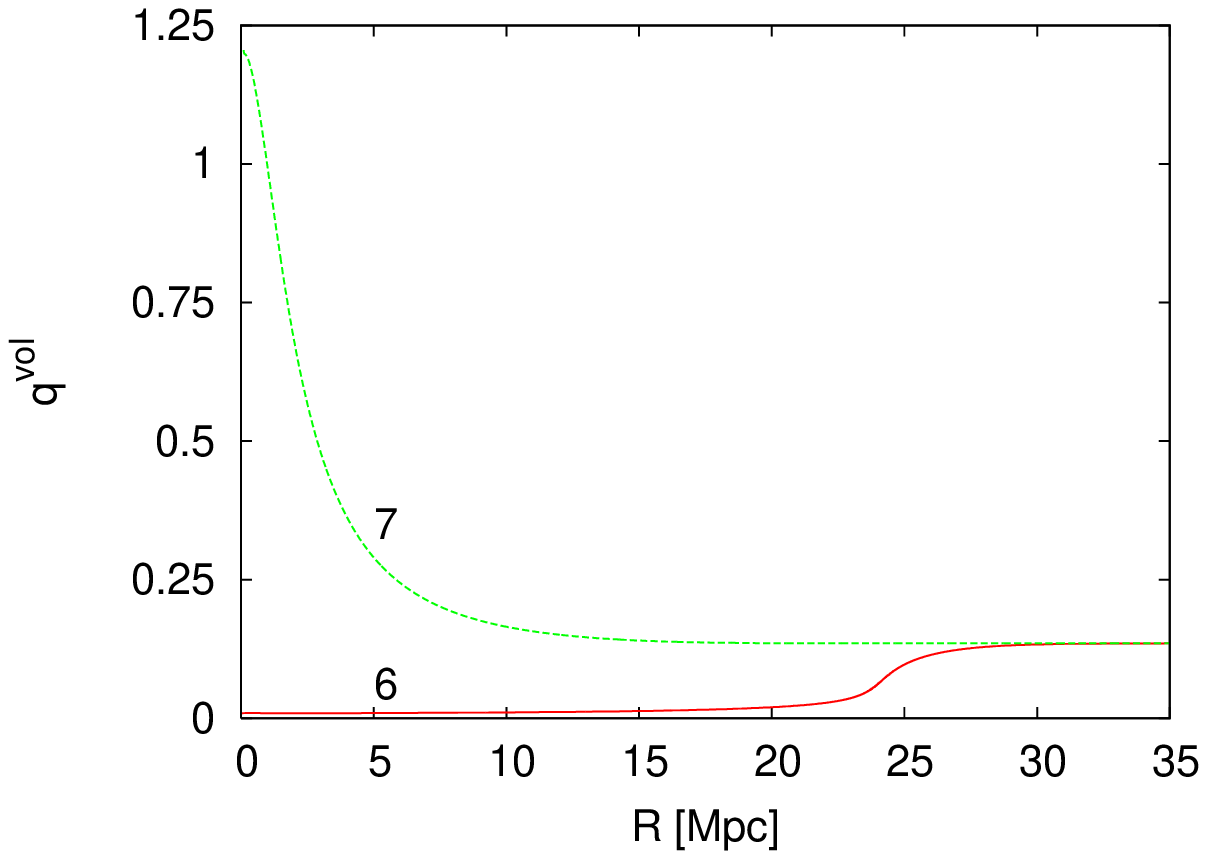}
\caption{The current density distribution (left panel) and volume deceleration
parameter (right panel) for models of cosmic structures (models 6, 7).}
\label{f4}
\end{figure}

\begin{figure}
\includegraphics[scale=0.65]{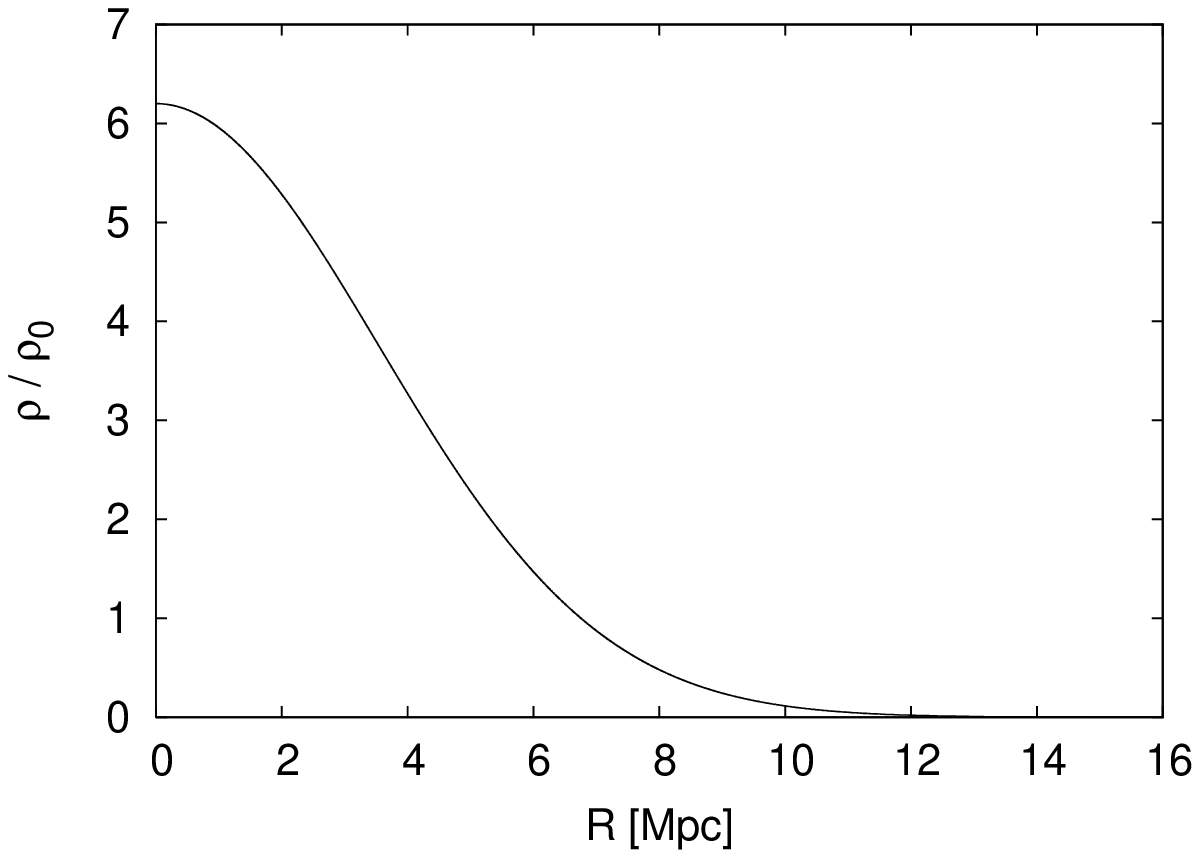}  
\includegraphics[scale=0.65]{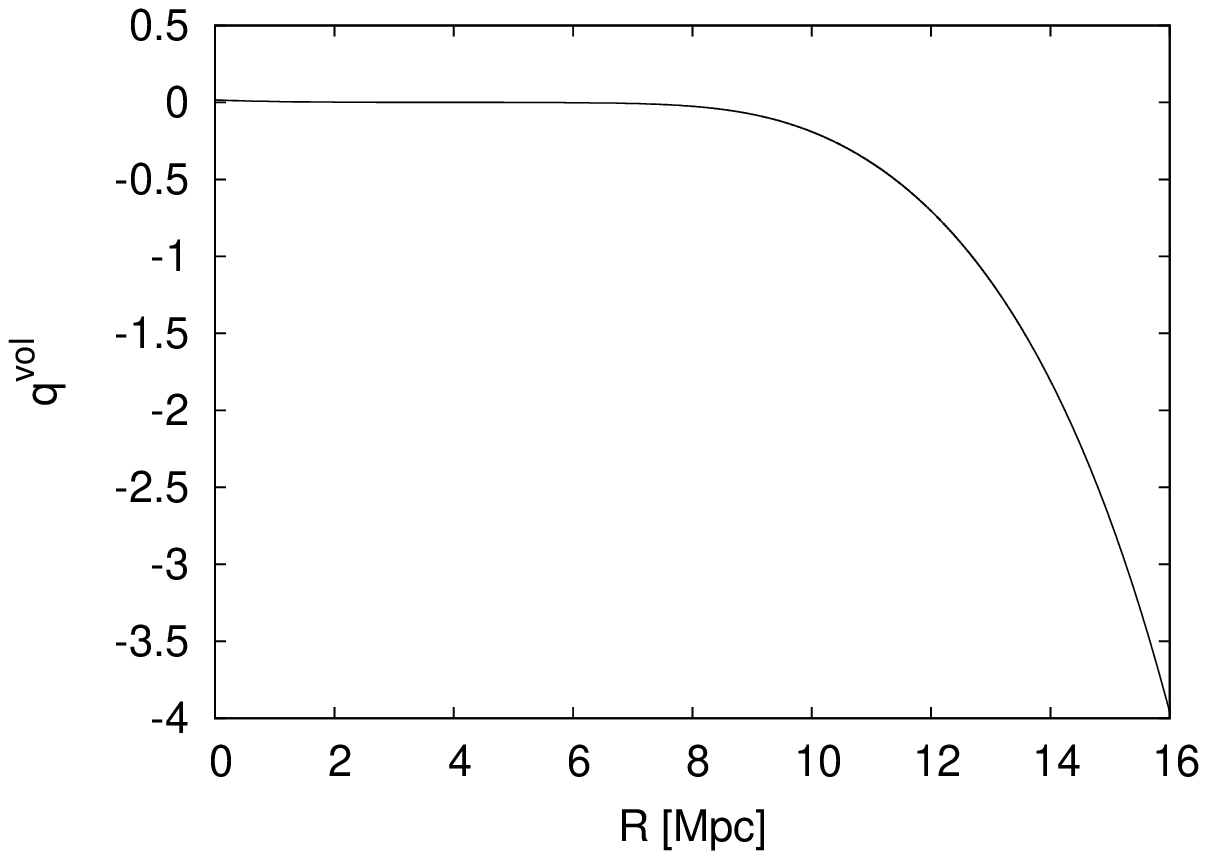}  
\caption{The current density distribution (left panel) and
deceleration parameter (right panel) for model 8.}
\label{f5}
\end{figure}

Now let us examine the volume deceleration parameter within models of cosmic
voids and superclusters.  Figure \ref{f4} presents density distribution of
realistically evolving cosmic structures (void -- curve 6, supercluster -- curve 7). 
It can be seen from the right panel of 
figure \ref{f4} that the 
volume deceleration parameter within these models is positive.
As above, we can modify our models in such a way that the volume
deceleration parameter is negative, but again this leads to a very large amplitude of
$t_B$. For example, in model 8 whose density and the volume deceleration parameter are presented in figure \ref{f5}\footnote{
Employing a model of qualitatively similar features as model 8, Hossain \cite{H07}
showed that the observer situated at the origin 
in order to successfully employ the Friedmann model has to assume the existence
of dark energy.} the volume deceleration parameter is negative. 
However, the bang time function in model 8 is of amplitude $\approx 11 \times 10^9$ y,
which leads to unrealistically small age of the Universe.

\section{Conclusions}\label{concl}
In this paper we have studied 
the relation between the volume deceleration parameter
obtained within the Buchert averaging scheme and the deceleration parameter
derived from the observations of supernovae. 
This work was motivated by recent
results showing there there are models which despite $\Lambda=0$
and average expansion rate is accelerating, i.e.
$\ddot{a} > 0$ [where $a$ is defined by relation (\ref{aave})].
This opens the
possibility that backreaction and averaging effects may be used 
as an interesting alternative explanation to the dark
energy phenomenon.  

We have compared the quantities obtained within the exact and
inhomogeneous models with their average counterparts. 
We focused on the
supernova observations and models of cosmic structures. For this purpose the
Lema\^itre--Tolman model was employed. It was showed numerically that the averaging of
models which fit the supernova observations does not lead to volume
acceleration ($\ddot{a} < 0$ for these 
averaged models and hence $q^{vol} >0$).
It was also shown that
realistically evolving models of cosmic structures have also $q^{vol} >0$.
It was possible to modify these model in such a way that after the
averaging $q^{vol} <0$. This was obtained by choosing $E$ function of positive amplitude
- as was recently proved by Sussman \cite{S08} this is a necessary condition to obtain 
$q^{vol} <0$. However, in models with realistic density distribution,
in such a cases, $E \gg 1 \gg M/R \approx 10^{-7} - 10^{-6}$, hence as seen from (\ref{evo}) $t_B \approx t$
(to remind $c \times 10^{10}$y $\approx 3$ Gpc). Thus, within such models
the age of the Universe is unrealistically small.

Our analysis has been performed in the limited class of Lema\^itre--Tolman models, which due to their spherical symmetry are arguably too
simple to give a full understanding of averaging and backreaction problems. However, within this class, we conclude that the volume  deceleration parameter $q^{vol}$ 
is not a quantity which can be directly related to observations.

It is possible that the volume deceleration parameter $q^{vol}$ becomes
negative only after averaging over the scales which are larger than 100 Mpc. On such
large scales the structure of the Universe becomes too complicated to be
fully described by spherically symmetric models. However,
it is intriguing that models which fit the supernova observations
 and for which the distance deceleration parameter, $q^{dis}$,
is negative have still $q^{vol}>0$. This suggest
that the volume  deceleration $q^{vol}$ 
does not have a clear interpretation in terms of observable quantities. 
It does not, of course, mean that averaging and backreaction effects 
cannot potentially be employed to explain the phenomenon of dark energy. 
However, our work here indicates that such a 
potential solution of the dark energy problem should be based upon different
methods than those related to volume deceleration parameter.
Rather than showing that $q^{vol}<0$ the averaging approach should explain
observations -- reproduce correct values of distance
to supernovae, correct shape of the CMB power spectrum, etc.
An interesting,  quasi-Friedmannian approach, was recently suggested in \cite{LABKC08}.
In this approach backreaction is modeled in terms of 
the morphon field \cite{BLA06}.
In such a case a Universe is describe by a homogeneous model with the spatial
curvature being just a function of time. As shown in \cite{LABKC08}
such approach lead to an agreement with supernova and CMB data
without the need for dark energy, but requires $q^{vol}<0$.

\ack

We would like to thank Henk van Elst and Thomas Buchert for useful discussions and comments.
KB would like to thank Peter and Patricia Gruber 
and the International Astronomical Union for the PPGF Fellowship,
the support of the Polish Astroparticle Network (621/E-78/SN-0068/2007)
is also acknowledged.
KB is also grateful to LA and the Albert Einstein Institute, where part of this research was carried out, for their hospitality. 
LA acknowledges the support of the NSF with grants 
DMS-0407732 and DMS-0707306 to the University of Miami.

\appendix

\section{Model specification}

\setcounter{section}{1}

\begin{figure}
\includegraphics[scale=0.65]{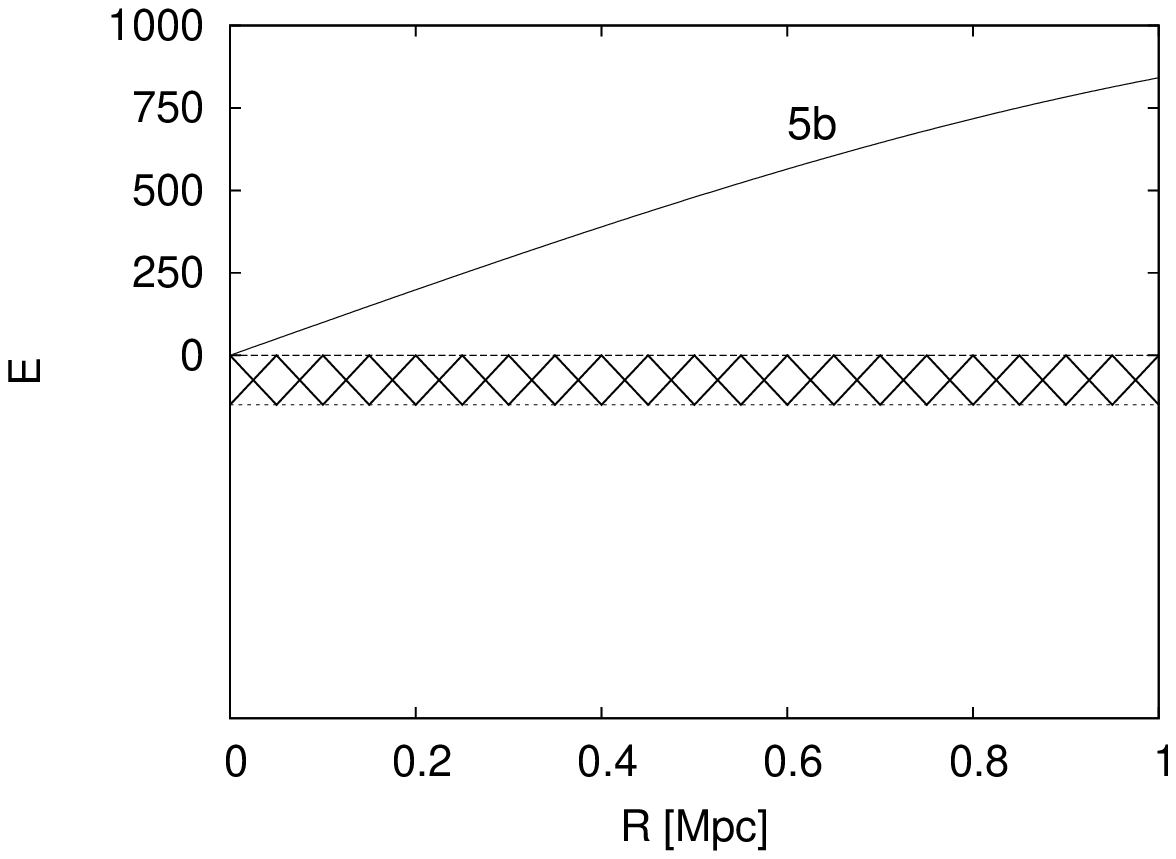}  
\includegraphics[scale=0.65]{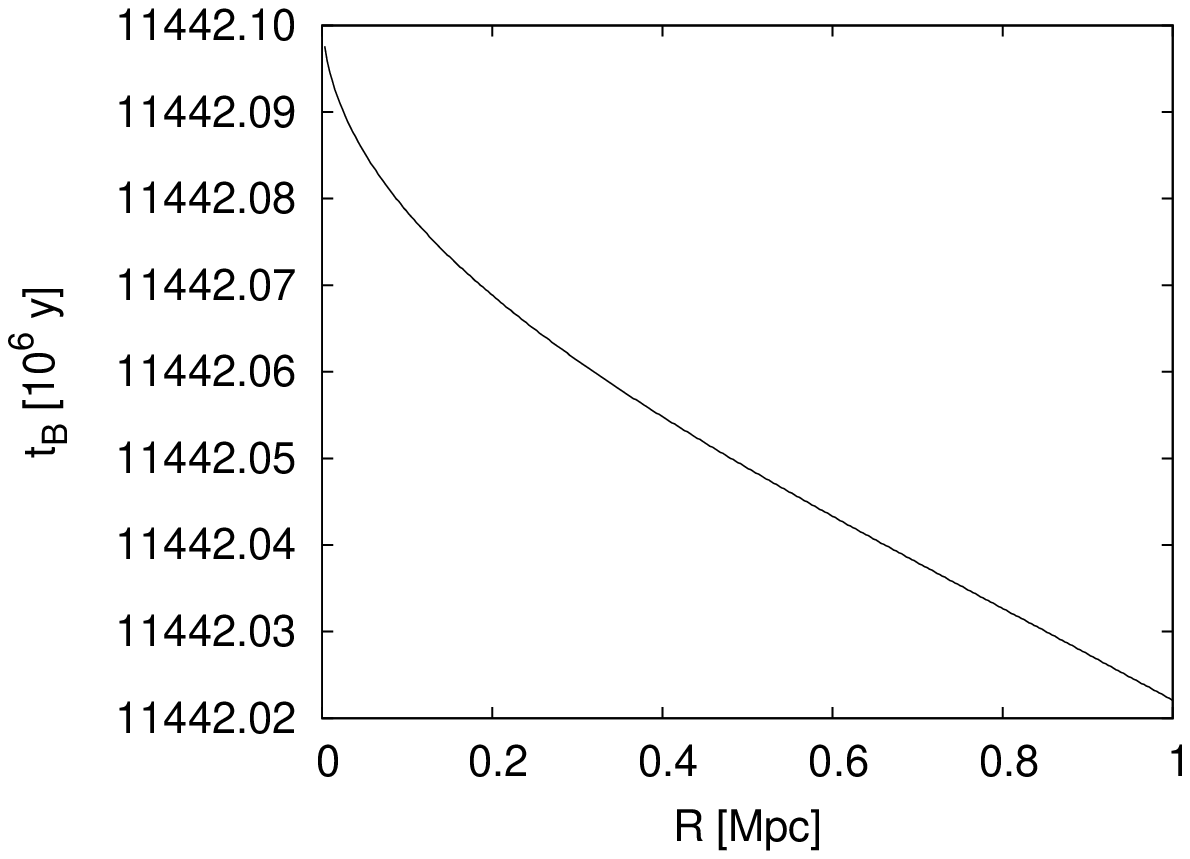}  
\caption{Left panel presents the function $E(r)$  for models 5a and 5b.
Please note that the y-scale in the upper part of the left panel is
different than in the lower part. Right panel bang time function 
for model 5b.}
\label{figA1}
\end{figure}

\begin{figure}
\includegraphics[scale=0.65]{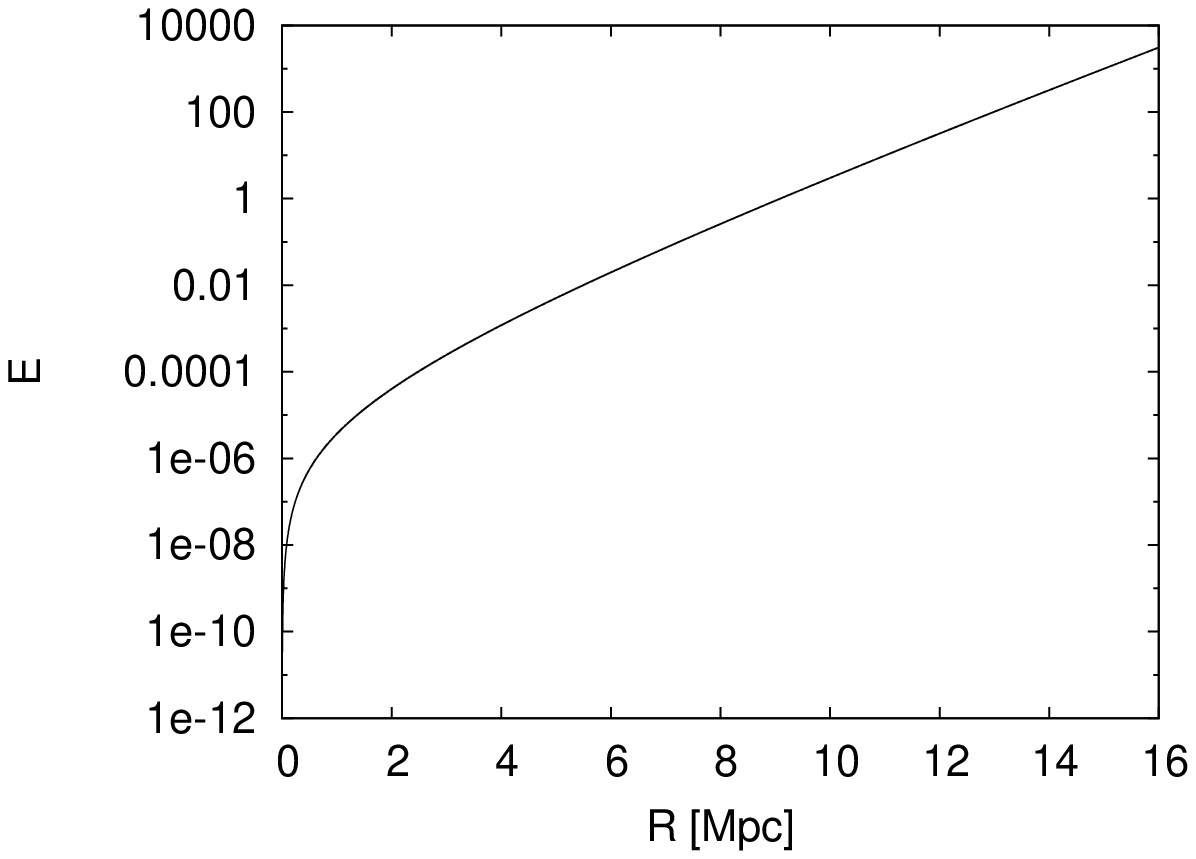}  
\includegraphics[scale=0.65]{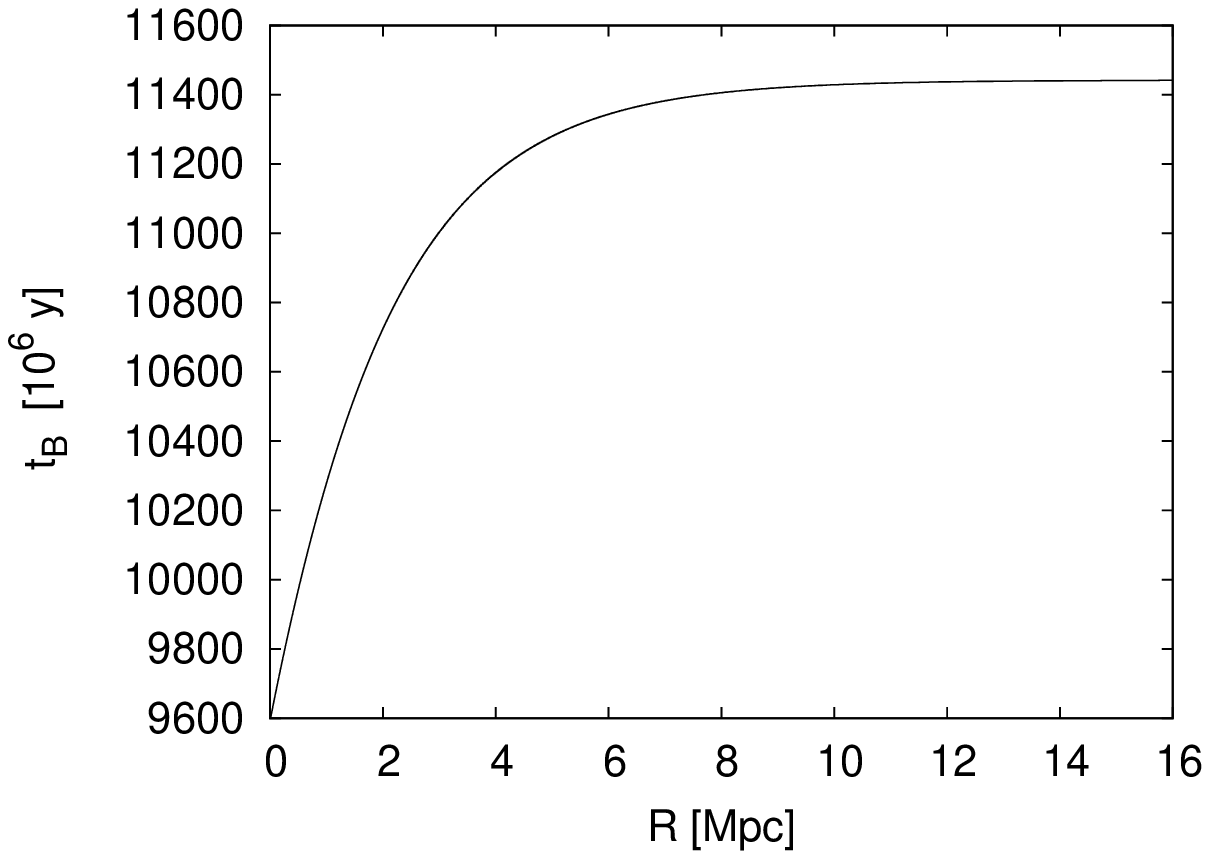}  
\caption{The function $E(r)$ (left panel) and $t_B(r)$ (right panel) for model 8.}
\label{figA2}
\end{figure}

There are three arbitrary functions of the radial coordinate  
in the Lema\^itre--Tolman. However only two functions are 
independent and the third one is specified by the choice of the
radial coordinate. Models considered in this paper are defined as follows:

\begin{enumerate}

\item
Model 1 and 2

The radial coordinate is chosen as the present day value of the
areal distance $r:=R_0$. Models 1 and 2 are specified
by the present day density distribution and the time bang function.
The density distribution is parametrized by

\begin{equation}
\rho(t_0,r) = \rho_b \left[ 1 + \delta_{\rho} - \delta_{\rho} \exp \left( - \frac{r^2}{\sigma^2} \right) \right],
\label{rhofl}
\end{equation}
where $\rho_b = \Omega_m \times  (3H_0^2)/(8\pi G)$, $\Omega_m = 0.27$, $H_0 = 70$ km s$^{-1}$ Mpc$^{-1}$.
In model 1 $\rho_{\delta} = 1.9$, $\sigma = 0.9$ Mpc, and 
in model 2 $\rho_{\delta} = 1.5$, $\sigma = 0.5$ Mpc.
In these models the big bang is assumed to occur simultaneously at every point, i.e. $t_B = 0$.
The functions $M$ and $E$ are then calculated and using 
eqs. (\ref{den}) and (\ref{evo}) respectively.

The time instants as well as background density $\rho_b$ 
in all models (1--8) is chosen as density of a Friedmann model ($\Omega_m = 0.27$, $H_0 = 70$ km s$^{-1}$)
and time instants are calculated using the following formula \cite{P93}:

\begin{equation}
t(z) =  \frac{1}{H_0} \int\limits_{z}^{\infty} \frac{d \tilde{z}}{(1+\tilde{z}) \sqrt{ \Omega_{mat} (1+\tilde{z})^3 + \Omega_K (1+\tilde{z})^2} },
\label{tz}
\end{equation}
where $\Omega_K = 1 - \Omega_{m}$.
The last scattering instant ($t_{LS}$) is set to take place when $z = 1089$ and the current instant ($t_{0}$) when $z=0$ --- $t_{LS} = 4.98 \times 10^5$ y, and $t_0 = 11.4421 \times 10^9$ y.

\item
Model 3 and 4

As above the radial coordinate is chosen  as the present day value of the areal distance $r:=R_0$. These two models are defined by the current expansion rate, 
and assumption that $\rho(t_0,r) = \rho_b$. The expansion rate is parametrized using

\begin{equation}
H_T(t_0,r) = \frac{\dot{R}}{R} = H_0 \left[ 1 - \delta_{H} + \delta_{H} \exp \left( - \frac{r^2}{\sigma^2} \right) \right],
\label{expfl}
\end{equation}
where $H_0 \delta_{H} = 9.6$ km s$^{-1}$ Mpc$^{-1}$, $\sigma =0.6$ Mpc,
and $H_0 \delta_{H} = 12$ km s$^{-1}$ Mpc$^{-1}$, $\sigma =1.2$ Mpc
for model 3 and 4 respectively.
In these models density is assumed to be homogeneous at the current epoch.
The function $M$ is then calculated using the above relation and eq. (\ref{vel}).
It should be noted that the $H_T$ is one of several generalization 
of the Hubble constant, which in the Friedmann model
is $H_0 = \dot{a}/a$. Apart from the transverse Hubble parameter, $H_T$, one can also define the radial  Hubble parameter, $H_R$, [see eq. (\ref{qddf})],
and the volume Hubble parameter defined as $H_V = (1/3) \Theta = H_R + 2 H_T$.

\item
Model 5a

The radial coordinate is chosen as the present day value of the areal distance, i.e. $r:=R_0$.
The model is defined by density distributions given at the present instant
and at last scattering.
The density distribution at the current instant is parametrized by

\begin{equation}
 \rho(t_0,r) = \rho_{b} \frac{\delta}{(r/r_s)(1+r/r_s)^2} ,
\label{nfwp}
\end{equation}
where $\delta = 28 170$ and $r_s = 191 kpc$. This is a 
Navarro, Frenk, and White galaxy cluster profile \cite{NFW}.
As can be seen this profile is singular at the origin but
this problem can be overcome by matching the NFW profile
with a singular--free profile as $f(r) = - a r^2 + b$.

The density profile at last scattering is assumed to be homogeneous, thus the areal distance at last scattering is:

\begin{equation}
R (t_{LS},r) = \left( \frac{M}{\kappa \rho_{LS} c^2} \right)^{1/3}.
\end{equation}
The function $M(r)$ is then calculated from eq. (\ref{den}).
Function $E$ can be calculated by subtracting solutions
of (\ref{evo}) for $t_{LS}$ and $t_0$ (for details see \cite{KH02}).
The function $E$ is presented in the left panel of figure \ref{figA1}.

\item 
Model 5b 

The radial coordinate is chosen as the present day value of the
areal distance, $r:=R_0$.
The model is defined by density distribution given by (\ref{nfwp})
and $E$ of the following form

\begin{equation}
E (r) = 10^3 \sin \left(10^{-3} r {\rm Mpc}^{-1} \right).
\end{equation}
This profile is presented in the left panel of figure \ref{figA1}
and the bang time function $t_B$ in the right panel.

\item
Models 6 and 7

The radial coordinate is chosen as the value of the
areal distance at last scattering instant, $r:=R_{LS}$.
Model 6 and 7 are defined by the assumption that $t_B=0$
and the density distribution,
which at last scattering is of the following form

\begin{equation}
\rho(t_{LS},r) = \rho_b \left( 1 - \delta \exp{(-a \ell^2 r^2)}
+ \gamma  \exp{\left[-\left(\frac{\ell r - c}{d}\right)^2 \right]} \right), 
\end{equation}
where  $\ell = 1/kpc$;
$\delta = 1.2 \times 10^{-3}$ and $2 \times 10^{-3}$
for model 6 and 7 respectively;
$\gamma = 14.62 \times 10^{-4}$ and $8.03 \times 10^{-4}$
for model 6 and 7 respectively;
$a = 0.01$ and $0.04$ for model 6 and 7 respectively;
$c = 18$ and $12$ for model 6 and 7 respectively;
and $d=6$ and $5$ for model 6 and 7 respectively.
The bang time function for both these models is $t_B = 0$.
The mass function, $M(r)$ is calculated
from eq. (\ref{den}), and the function $E(r)$ is calculated from eq. (\ref{evo}).

\item
Model 8

The radial coordinate is chosen  as a present day value of the
areal distance: $r:=R_0$. Density distribution is of the following
form

\begin{equation}
\rho(t_{0},r) = 6.2 \rho_b \exp{(-4 \times 10^{-8} (\ell r)^2)}, 
\end{equation}
and the function $E$ is

\begin{equation}
E(r) =  \left(\frac{H_0}{c} r \right)^2 \exp{(10^{-3} \ell r)}, 
\end{equation}
which except for $[\exp{(10^{-3} \ell r)}]$ is the same 
as $E(r)$ profile in the empty Universe.
This profile is presented in the left panel of figure \ref{figA2}
and the bang time function $t_B$ in the right panel.

\end{enumerate}

\end{document}